\documentclass[fleqn,twoside]{article}
\usepackage{espcrc2}
\pdfoutput=1

\usepackage{graphicx}

\newcommand{\AmS}{{\protect\the\textfont2
  A\kern-.1667em\lower.5ex\hbox{M}\kern-.125emS}}

\title{IceCube: Status and First Results}

\author{P. Berghaus\address[UW]{University of Wisconsin, Madison (IceCube Project); Madison, WI 53703; USA} for the IceCube Collaboration\thanks{www.icecube.wisc.edu/collaboration/authorlists/2008/4.html}}
       
\begin{document}

\begin{abstract}
IceCube is a cubic neutrino telescope under construction at the South Pole since the austral summer 2004/2005 with a total instrumented volume of the order of 1 ${\rm km}^{3}$. At the moment it is taking data with 40 deployed strings. The full detector is expected to be completed in 2011 with up to 80 strings holding 60 digital optical modules (DOMs) each. The progenitor detector AMANDA has been operating at the same site since 1997 and is still functioning as a means to enhance neutrino effective area at energies below 100 GeV. A summary of science results and status of the project is presented.
\vspace{1pc}
\end{abstract}

\maketitle

\section{INTRODUCTION}
\subsection{High-Energy Neutrino Astrophysics}

Neglecting gravity, neutrinos interact only by way of the weak interaction. This precludes absorption by matter and radiation fields which affect photons and therefore potentially allows observation of objects and processes that are not accessible to conventional gamma-ray astronomy. Targets of interest include the neighborhood of black holes and active galaxies, where photonic emissions are subject to absorption by ambient matter and extragalactic background light. The obvious disadvantage resulting from the restriction to weak interactions is the need for very large detector volumes to achieve at least a moderate event yield. 

One of the longest-standing problems in astrophysics is the search for the origin of cosmic radiation \cite{Halzen:2002pg}. Since the direction of charged cosmic rays, except at the very highest energies, effectively gets scrambled by magnetic fields, any information about their origin must come from electrically neutral particles, i.e. photons and neutrinos. Investigation of this issue is one of the main goals of Very High Energy (VHE) neutrino detectors, with instrumented volumes up to one cubic kilometer. These take advantage of Cherenkov radiation of charged particles produced in interactions of neutrinos with ambient matter. The large fiducial volume required for the detection of the comparatively low neutrino fluxes at TeV-PeV energies makes it necessary to take advantage of large natural occurrences of light-transparent media, such as the deep sea or polar ice sheets.

So far, no VHE neutrino signal from outside the solar system has been detected. This situation is expected to change after completion of a new generation of detectors with an instrumented volume of the order of 1 ${\rm km}^3$ \cite{Halzen:2007ip}.

\begin{figure}[htb]
\includegraphics{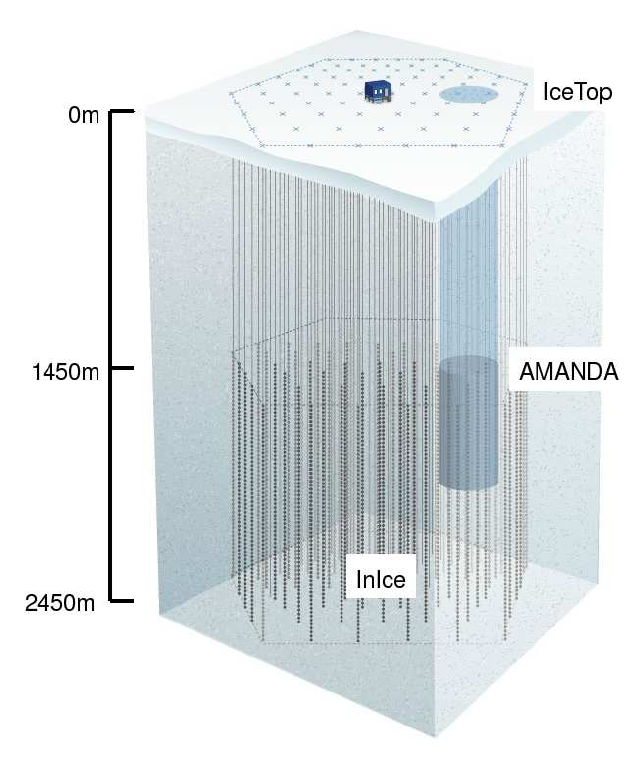}
\caption{Schematic view of the IceCube detector, showing the three currently operating components: IceTop surface air shower array, AMANDA detector and InIce strings. As of 2008, IceTop and InIce are both completed to 50\% (40/80).}
\label{fig:icecube_scheme}
\end{figure}

\section{THE ICECUBE DETECTOR}

\subsection{Overview}

IceCube is expected to become the first operational cubic-kilometer scale neutrino detector. Built at the geographic South Pole around its predecessor AMANDA, the sub-surface detector (InIce) is scheduled to reach its full extent of 80 vertical strings carrying 60 digital optical modules (DOMs) each by the year 2011 \cite{Hill:2006mk}. Already, with 40 deployed strings, it is the largest neutrino telescope ever built and has started delivering data of unprecedented quality. An additional detector component, the surface air-shower array IceTop, will eventually comprise 80 pairs of frozen water tanks with two DOMs in each. Figure \ref{fig:icecube_scheme} shows a schematic view of the three components of IceCube in its current configuration.

While the main focus of IceCube is astrophysics, there is also a wide range of other research topics. These include investigation of neutrino oscillations, indirect searches for dark matter and exotic particle physics, such as direct searches for magnetic monopoles. Due to its location, IceCube will be sensitive mostly to neutrino signals from the northern hemisphere, since above the horizon background from downgoing muons makes separation of the neutrino signal very challenging.

\subsection{Hardware}

So far, 2400 DOMs have been deployed, with an overall failure rate of about 2\% \cite{little_spencer}. Each DOM contains a photomultiplier tube facing downward and two boards for readout and calibration electronics. Waveforms recorded by the photomultiplier are digitized {\it in situ} and transmitted to the surface by twisted-pair cables located inside the strings. Surface readout electronics are situated inside a counting house located at the center of the array.

To account for the various event topologies, several diferent hardware triggers were implemented in IceCube. The most important is the Single Majority Trigger (SMT8), requiring a signal corresponding to about 0.3 photoelectrons (``hit'') in eight DOMs within a time window of 5000 ns. Only those hits are counted for which a local coincidence requirement is fulfilled, meaning that at least one neighboring or next-to-neighboring DOM must also register a signal within 1000 ns. Data falling within $\pm 10\mu s$ of the trigger time are combined to separate events. Several simple reconstruction algorithms are then applied in real time and those events that pass a set of simple quality cuts are transmitted to the northern hemisphere by satellite. About 6\% of all events are selected in this way, corresponding to a data rate of slightly more than 30 GB/day.

\subsection{Status}
\begin{figure}
\includegraphics[width=3in]{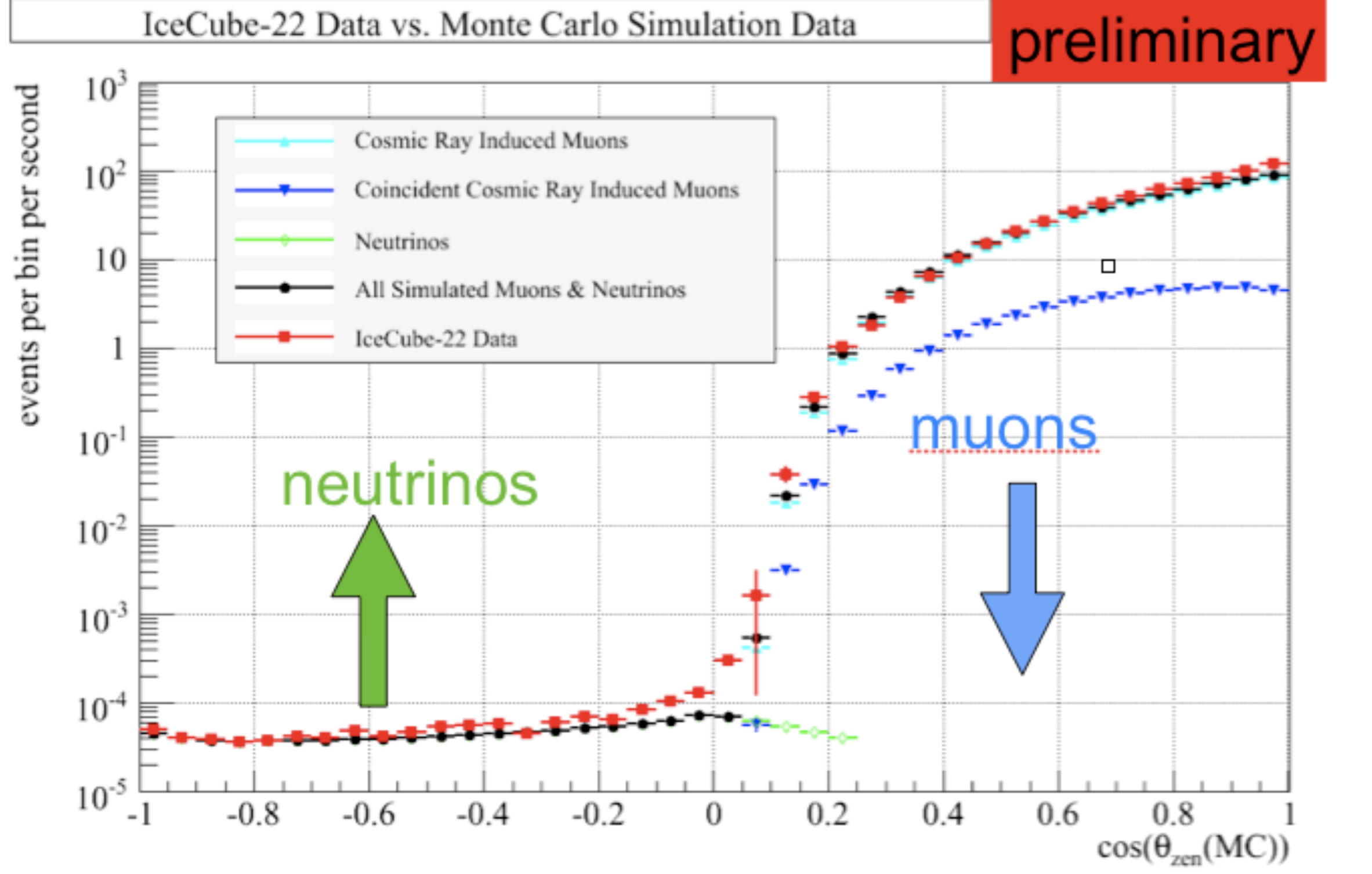}
\caption{All-Sky Muon Flux as measured with IceCube-22, renormalized to trigger level. The data in this plot represent a total livetime of 275.6 days. The downgoing muon flux has been simulated using CORSIKA/SIBYLL and is shown separately for single and multiple-primary (coincident) cosmic ray showers. Muon neutrino events were only simulated below a zenith angle of 70 degrees.}
\label{fraeuleinrosenauplot}
\end{figure}

\begin{figure}
\includegraphics[width=3in]{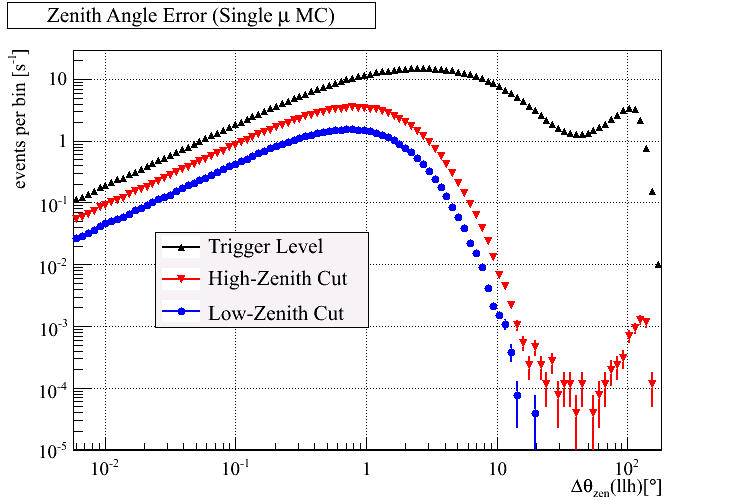}
\caption{Zenith angle error for muon tracks depending on quality cut level. The ``High-Zenith'' Cut as used to generate the component of Fig. \ref{fraeuleinrosenauplot} with $cos\theta>0.2$ shows a residual tail of events that have no correlation to the true direction. After the tighter ``Low-Zenith'' cut, applied only to events around and below the horizon, the tail has vanished. The median resolution varies from 0.7 to 0.8 degrees depending on the zenith angle.}
\label{delta_zen}
\end{figure}

After the 2007/08 austral summer deployment, a total of 40 strings have been completed. The SMT8 trigger rate in this configuration is about 1000 Hz, which is expected to rise to 1650 Hz in the full 80-string detector \cite{all_bracked}. In addition to the InIce strings, 40 IceTop surface tanks are operating, each containg two DOMs. Construction of the detector is proceeding according to schedule and is should be completed by 2011. Fig. \ref{fraeuleinrosenauplot} shows the all-sky muon flux as measured with 2007 data using 22 InIce strings. The result demonstrates that simulation of the detector response is reasonably well understood for track-like events coming from all zenith angles. The simulated angular resolution at trigger level and after two different quality cuts is shown in Fig. \ref{delta_zen}.

\section{PHYSICS RESULTS}

\subsection{Astrophysical Point Sources}

Processes leading to the production of VHE $\gamma$-radiation can be subdivided into two classes. In \emph{electronic} emission mechanisms, gamma rays are produced by inverse Compton scattering of high energy electrons on ambient photons. Conversely, \emph{hadronic} models assume gamma production through the decay of neutral pions, which in turn are produced in interactions involving high-energy protons. Observations by gamma-ray telescopes have so far not provided unambiguous evidence for either model, irrespective of the source type \cite{Aharonian:2007bn}.

Neutrino astrophysics provides a natural means to distinguish between the two mechanisms, since hadronic interactions will also produce charged mesons, all of which decay into final states containing at least one neutrino. Purely electronic emission processes produce neutrinos only in negligible numbers. Any detection of neutrinos from a specific source would therefore constitute very strong evidence for the hadronic model \cite{Halzen:2007ip} and consequently allow to identify the sources of cosmic radiation.
The search for point sources of extraterrestrial neutrinos is a major analysis topic in IceCube. In order to improve sensitivity, various sophisticated search methods have been developed \cite{Braun:2008bg}. 

\begin{figure}
\includegraphics[width=2.63in]{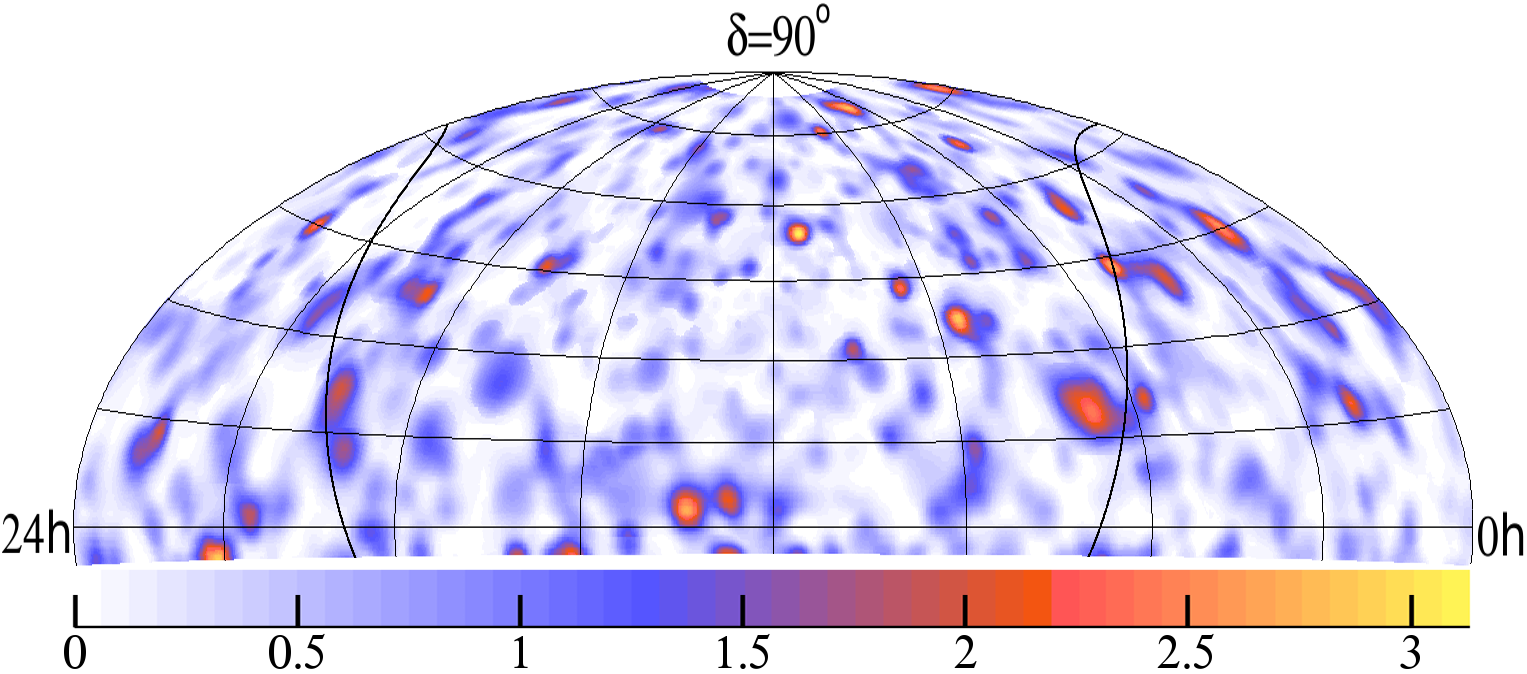}
\includegraphics[width=3in]{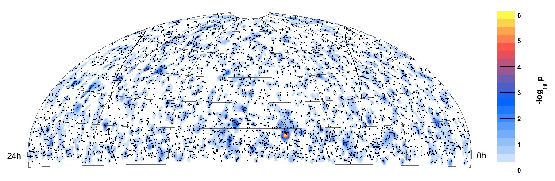}
\caption{Significance maps for neutrino point source searches. Top: Result from 2000-2006 AMANDA analysis. Bottom: IC22 unbinned search. Note the different scales on each plot. With respect to the AMANDA result, the average sensitivity in the IC22 analysis has improved by a factor of two.}
\label{ps_plots}
\end{figure}

A final result for the AMANDA, covering seven years of detector operation, has recently been completed \cite{ps_amanda}. The result, along with that for IceCube-22 \cite{jl_now}, is shown in Fig. \ref{ps_plots}. Both the AMANDA and IceCube data are fully consistent with background fluctuations. The IceCube result shows a slight feature at 11 degrees below the horizon. The background probability, taking into account all trial factors, was determined to be 1.3\%, corresponding to approximately $2.3\sigma$. Data taken during the 2008 season using 40 strings should be sufficient to make a definite statement about the nature of this excess.

\subsection{Diffuse Neutrino Flux}
The diffuse analysis represents an alternative approach in the search for an astrophysical neutrino signal. Here, an all-sky search is conducted looking for an excess at high neutrino energies. To separate the signal from background, an energy-correlated parameter cut is applied after carefully cleaning the data of downgoing muon tracks. For AMANDA data from 2000 to 2003, this method yielded 6 events in the final sample, compared to an expected background of 6.1. This corresponds to a limit for diffuse muon neutrino flux from the northern hemisphere of $E^2\Phi<7.4\times10^{-8}{\rm GeVcm}^{-2}{\rm s}^{-1}{\rm sr}^{-1}$ for an assumed $E^{-2}$ spectrum with $15.8$ TeV$<E_\nu<2.5$ PeV, covering 90\% of the simulated signal \cite{diffuse}. The result also has implications for prompt (charm) production of atmospheric neutrinos, ruling out some of the highest-yielding models at 90\% C.L.

\subsection{Other Physics Analyses}

Searches for neutralino WIMP signals from the Sun and Earth have been completed using data from 2001 and 2001-2003 respectively \cite{Achterberg:2006jf}. The corresponding result for IC22 will be published soon. The limits on the muon flux from neutralino annihilation are consistent with or slightly better than those from other indirect searches, in spite of AMANDA's shorter integrated live time. Analyses of the remainder of AMANDA data are ongoing \cite{daan}.

Gamma Ray Bursts (GRB) are believed to be a likely source for ultra-high energy cosmic rays and high energy neutrinos, motivating a search for a neutrino signal in coincidence with transient gamma signals from satellite-borne detectors. For the years 2000-2003, 407 bursts were analyzed looking for corresponding up-going muon tracks, but yielding no neutrino candidate events inside the temporal and spatial search windows. Also, ``rolling'' searches were conducted looking for an excess of cascade events within a predefined time independent of known burst alerts. All results were consistent with background expectation. The best limit so far, obtained by the coindicent muon analysis, lies slightly above the Waxman-Bahcall flux, one of the standard benchmarks in calculating neutrino yields from GRB \cite{grbpaper}. At the time of the exceptionally bright GRB 080319B \cite{Racusin:2008pd}, IceCube was taking data with 9 strings. A search for neutrino emission has been conducted and will be published soon.

Further studies have been undertaken looking for Ultra-High Energy (UHE) neutrinos \cite{lisa_gerhardt}, magnetic monopoles \cite{hickes} and diffuse neutrino fluxes based on detection of electromagnetic and hadronic cascades from $\nu_e$, $\nu_{\tau}$ and neutral current interactions \cite{oksana}. None of the results showed a statistically significant excess over background.

Remarkably, the first extraterrestrial object detected by IceCube was the sun, seen by the IceTop surface array. During a solar flare in 2006, the energetic particle spectrum was measured in the energy range 0.6-7.6 GeV through variations in the discriminator counting rate \cite{Abbasi:2008vr}.

\section{OUTLOOK}
As the IceCube detector increases in volume, new possibilities for physics analyses will arise. One entirely new path of investigation is the search for tau neutrino interactions. Due to flavor oscillations, one out of three of all astrophysical neutrinos arriving at the Earth is expected to be a $\nu_{\tau}$. For this type of interaction, several characteristic signatures have been identified. The major advantage in detecting tau neutrinos is the very low background from atmospheric interactions, making detection of an extraterrestrial signal significantly less ambiguous than in other channels (e, $\mu$) \cite{Cowen:2007ny}.

Operation of the AMANDA detector as a subsystem of IceCube will be concluded with the end of the 2008 season. Eventually, its function will be taken over by the six Deep Core strings, the first of which should be deployed during the austral summer 2008/09. On each of these strings, 50 DOMs will be arranged with a spacing of 7m starting at a depth of approximately 2100 m, which corresponds to a region of exceptionally clear ice \cite{lizzy_r}.

 In this low-energy extension, the 80 main IceCube strings can be used to veto tracks coming from above the horizon, expanding the active aperture for detection of muon neutrinos to the southern hemisphere. Since this is where the bulk of the galactic plane, and hence the majority of potential galactic neutrino sources, is located, it can be expected to have a significant impact on science operations. An additional effect of background vetoing is a decrease in the effective muon neutrino energy threshold to values of 50 GeV or even less \cite{big_andy}.

A variety of analyses will make use of the a combination of several IceCube components. The addition of the Deep Core will allow to set more stringent limits on solar WIMPs. Preliminary investigations indicate an improvement by up to an order of magnitude for low mass neutralinos \cite{gustav}. 

The IceTop surface array together with the InIce component should be able to study cosmic rays up to the EeV range. By combining surface radius and in-depth intensity of muon showers, IceCube will be able to make cosmic ray composition measurements, especially around the ``knee'' at $E_{primary}\approx3{\rm PeV}$ \cite{icetop}.

\section{CONCLUSION}

The transition from AMANDA to IceCube has so far been remarkably successful. First physics results have been published, validating the excellent performance of the new detector. As IceCube continues to grow with every deployment season, increases in both volume and angular resolution should soon allow results to significantly surpass those from AMANDA. An important symbolic milestone for IceCube was reached in 2008, as integrated exposure exceeded $1{\rm km}^3{\rm yr}$. Data becoming available over the course of the next few years will allow important physics investigations, such as a probe of the diffuse Waxman-Bahcall flux and exclusion or confirmation of various emission models for individual cosmic objects.

\end{document}